\documentclass[fleqn,usenatbib]{mnras}

\usepackage{newtxtext,newtxmath}
\usepackage[T1]{fontenc}
\usepackage{ae,aecompl}

\usepackage{graphicx}	
\usepackage{amsmath}
\usepackage{multirow}
\usepackage{lipsum}
\usepackage{wasysym}
\usepackage{color}
\usepackage{comment}
\usepackage[squaren, Gray, cdot]{SIunits}

\title[Magnetic topology of Proxima Centauri]{The large-scale magnetic field of Proxima Centauri near activity maximum}

\author[B. Klein et al.]{
 \parbox[h]{\textwidth}{
Baptiste Klein$^{1}$,\thanks{E-mail: baptiste.klein@irap.omp.eu}
Jean-Fran\c{c}ois Donati$^{1}$,
\'Elodie M. H\'ebrard$^{1}$,
Bonnie Zaire$^{1}$,\\
Colin P. Folsom$^{1}$,
Julien Morin$^{2}$,
Xavier Delfosse$^{3}$,
Xavier Bonfils$^{3}$
}
\\
\\
$^{1}$Universit\'e de Toulouse, CNRS, IRAP, 14 av. Belin, 31400 Toulouse, France\\
$^{2}$LUPM, Universit\'e de Montpellier, CNRS, Place Eug\`ene Bataillon, F-34095 Montpellier, France\\
$^{3}$CNRS, IPAG, Universit\'e Grenoble Alpes, 38000 Grenoble, France\\
}

\date{Accepted. Received in original form 2020 October 27}

\pubyear{2020}

\newcommand{\pr}{$P_{\rm{rot}}$}
\newcommand{\vs}{$v \sin i$}
\newcommand{\kms}{km\,s$^{-1}$}

\newcommand{\mass}{$M_{\rm{S}}$}
\newcommand{\msun}{M$_{\odot}$}
\newcommand{\rsun}{R$_{\odot}$}
\newcommand{\rs}{$R_{\rm{S}}$}
\newcommand{\bl}{B$_{\ell}$}

\newcommand{\crr}{$\chi_{\rm{r}}^{2}$}
\newcommand{\chs}{$\chi^{2}$}

\newcommand{\mdw}{M\,dwarf}

\newcommand{\fii}{$f_{\rm{I}}$}
\newcommand{\fv}{$f_{\rm{V}}$}

\newcommand{\avbv}{<$B_{\rm{V}}$>}
\newcommand{\avbi}{<$B_{\rm{I}}$>}

\newcommand{\faxi}{$f_{\rm{axi}}$}

\newcommand{\ra}{$R_{\rm{A}}$}
\newcommand{\lxb}{$\log L_{\rm{X}}/L_{\rm{bol}}$}

\begin{document}
\label{firstpage}
\pagerange{\pageref{firstpage}--\pageref{lastpage}}
\maketitle

\begin{abstract}
We report the detection of a large-scale magnetic field at the surface of the slowly-rotating fully-convective \mdw\ Proxima Centauri. Ten circular polarization spectra, collected from April to July 2017 with the HARPS-Pol spectropolarimeter, exhibit rotationally-modulated Zeeman signatures suggesting a stellar rotation period of 89.8\,$\pm$\,4.0\,d. Using Zeeman-Doppler Imaging, we invert the circular polarization spectra into a surface distribution of the large-scale magnetic field. We find that Proxima Cen hosts a large-scale magnetic field of typical strength 200\,G, whose topology is mainly poloidal, and moderately axisymmetric, featuring, in particular, a dipole component of 135\,G tilted at 51$^{\circ}$ to the rotation axis. The large-scale magnetic flux is roughly 3$\times$ smaller than the flux measured from the Zeeman broadening of unpolarized lines, which suggests that the underlying dynamo is efficient at generating a magnetic field at the largest spatial scales. Our observations occur $\sim$1\,yr after the maximum of the reported 7\,yr-activity cycle of Proxima Cen, which opens the door for the first long-term study of how the large-scale field evolves with the magnetic cycle in a fully-convective very-low-mass star. Finally, we find that Proxima Cen's habitable zone planet, Proxima-b, is likely orbiting outside the Alfv\`en surface, where no direct magnetic star-planet interactions occur.
\end{abstract}

%The slack fraction of axisymmetric field suggests that Proxima Cen's habitable zone planet, Proxima-b, undergoes a roughly constant mass flux throughout is orbit and is well-shielded from cosmic rays at the time of observations. Finally, we find that Proxima-b is likely orbiting outside the Alfv\`en surface, excluding thus direct magnetic interactions from the planet to the star. 


\section{Introduction}

%Late \mdw s are primary targets in the quest for Earth twins \citep{kasting1993}. Their small masses and radii as well as their close-in habitable zones (HZ) make the detection of potentially habitable planets easier than for their solar counterparts.
%As a result, 

Late \mdw s are primary targets in the quest for Earth twins \citep{kasting1993}. Their small masses and radii, as well as close-in habitable zones (HZ), make the detection of temperate Earth-like planets around them easier than around solar-like stars. As a result, the most favorable planets to further investigate habitability with forthcoming telescopes like the JWST and ELTs orbit stars with spectral type later than M4 \citep[e.g.,][]{berta_thompson2015,anglada2016,gillon2017,dittmann2017,astudillo-defru2017b}. However, these stars exhibit strong magnetic activity \citep[e.g.,][]{west2011}, whose manifestations, such as high-energy winds or frequent flaring events, are likely to affect the properties of the planets in their HZ. Therefore, understanding the processes underlying activity phenomena is a major prerequisite to study the conditions of habitability around low-mass stars \citep{lammer2007,france2016}.

Late \mdw s are fully-convective \citep[FC;][]{baraffe1998}. Their underlying dynamo processes remain mysterious, despite recent advances in the explanation of observations by numerical models \citep{yadav2015,yadav2016}. Spectropolarimetric observations of FC stars have revealed a bimodal distribution of their magnetic properties with either strong axisymmetric dipoles, or weaker non-axisymmetric complex fields \citep{donati2006a,morin2008,morin2010,kochukov2017}. The origin of this bimodality, tentatively explained by bistability in the dynamo process \citep{morin2011b,gastine2012,gastine2013}, or a single oscillatory dynamo process \citep{kitchatinov2014}, is still debated in the literature. However, most of the observational results currently available involve very active stars whose dynamo lies in the so-called saturated regime, i.e., on the plateau of the activity-Rossby number\footnote{The Rossby number is defined as the stellar rotation period normalized to the convective turnover time, set to $\sim$143$^{+31}_{-22}$\,d for Proxima Cen using \citet{wright2018} empirical relationship.} relationship \citep[typically, Rossby number $Ro$\,\la\,0.1 which corresponds to a rotation period of roughly 10\,d for a mid-M dwarf;][]{pizzolato2003,kiraga2007,shulyak2017,Astudillo-Defru2017,wright2018,see2019}. Hence the interest in observing the magnetic field of FC \mdw s in the unsaturated dynamo regime like Proxima Centauri (rotation period of $\sim$90\,d), whose large-scale magnetic properties are still unconstrained.

Spectropolarimetry is the best way to model the large-scale topologies of stellar magnetic fields, and thereby the underlying dynamo processes powering magnetic activity. At the photospheric level, magnetic fields produce circularly-polarized Zeeman signatures that can be inverted into a map of the large-scale magnetic field using Zeeman-Doppler imaging \citep[ZDI,][]{donati2009}. Spectropolarimetric monitoring of FC \mdw s is thus expected to bring new constraints on the nature of the dynamo processes operating in their interiors. Moreover, the large-scale magnetic geometry is an essential ingredient for modelling stellar winds, known to play a major role in star-planet interactions \citep[SPI,][]{vidotto2014b,strugarek2015}.

%The total fractions of axisymmetric and poloidal magnetic energies in the reconstructed topology (resp. \faxi\ and \fpol) are excellent proxies of the magnetic cycle for partly-convective stars \citep[e.g.,][Lehman et al., in prep]{fares2009,boro_saikia2016,boro_saikia2018}. Long term spectropolarimetric monitoring of FC \mdw s hosting a magnetic cycle is thus expected to bring crucial constraints on the nature of the dynamo processes operating in their interiors. Moreover, the large-scale magnetic geometry is a crucial ingredient in modelling stellar winds, notorious to play a major role in star-planet interactions \citep[SPI,][]{vidotto2014b,folsom2019}.

Our closest neighbor, Proxima Centauri, is an active FC M5.5\,dwarf whose dynamo lies in the unsaturated regime \citep[$Ro$\,=\,0.63, \lxb\,=\,-3.94;][]{wright2018}. Proxima Cen hosts a HZ Earth-mass planet, Proxima-b \citep{anglada2016}, whose habitability conditions have been widely studied in the literature \citep[e.g.,][]{ribas2016,turbet2016,barnes2016,meadows2018}. This planet orbits at less than 0.05~au from its host star, and its atmospheric and surface properties are likely affected by stellar magnetic activity \citep[e.g.,][]{garraffo2016,garcia_sage2017}. Moreover, the star exhibits a 7\,yr-photometric activity cycle \citep{suarez2016,wargelin2017}, interpreted as resulting from a $\alpha \Omega$-dynamo at work in its convective interior \citep{yadav2016}. Spectropolarimetric observations of Proxima Cen can guide theoretical dynamo models and provide key inputs to investigate SPIs with the close-in planet.

In this study, we present the first reconstruction of Proxima Cen's large-scale magnetic topology from a set of 10 spectropolarimetric observations collected with HARPS-Pol from April to July 2017, around the activity maximum. In Sec.~\ref{sec:section2}, we present our spectropolarimetric observations and detection of magnetic field. We then detail, in Sec.~\ref{sec:section3}, the modelling of the Zeeman signatures and their inversion into the large-scale magnetic topology of Proxima Cen. Finally, we discuss the implications of our results regarding the activity and magnetic cycle of the star, and its extended magnetosphere, in Sec.~\ref{sec:section4}.

\section{Observations}\label{sec:section2}

\newcommand{\hd}{\hphantom{0}}
\begin{table}
    \centering
\caption{List of HARPS-Pol spectropolarimetrc observations of Proxima Cen (ESO program 099.C-0334(A), PI: H\'ebrard). All polarization sequences consist of 4 individual subexposures of 1200\,s each. Columns 1 to 3 indicate the observation dates, BJDs (at mid-exposure), and peaks of S/N (per 0.85\,\kms\ velocity bin), respectively. Columns 4 and 5 list the rms noise level relative to the unpolarized continuum level per pixel bin in each Stokes $V$ spectrum and the estimated longitudinal magnetic field \bl\ (with 1$\sigma$ error bars). The stellar rotation phase, computed from the reference time BJD\,=\,2457862 and a rotation period of 89.8~d, is given in column~6.}
\label{tab:obs_prop}
\begin{tabular}{ccccr@{$\pm$}lc}
\hline
Date & BJD & S/N & $\sigma_{\rm{V}}$ & \multicolumn2c{\bl} & Phase \\
 &  [2457000+]  & & [$10^{-4} I_{\rm{C}}$] & \multicolumn2c{[G]} & \\
\hline
2017/04/18 & 862.7525 & 130 & 4.5 & 3.2\, & \,8.2 & 0.008\\
2017/04/28 & 872.7921 & 154 & 3.8 & -27.7\, & \,6.6 & 0.120\\
2017/05/04 & 878.7772 & 157 & 3.7 & -73.5\, & \,6.9 & 0.187\\
2017/05/13 & 887.8047 & 78 & 7.0 & -36.7\, & \,18.8 & 0.287\\
2017/05/15 & 889.8118 & 91 & 6.4 & -52.0\, & \,15.1 & 0.310\\
2017/06/18 & 923.6802 & 105 & 5.4 & 5.8\, & \,12.5 & 0.687\\
2017/06/29 & 934.5817 & 96 & 5.7 & 12.7\, & \,14.9 & 0.808\\
2017/07/02 & 937.6045 & 145 & 4.0 & 5.2\, & \,8.0 & 0.842\\
2017/07/09 & 944.6375 & 165 & 3.5 & 19.4\, & \,6.7 & 0.920\\
2017/07/14 & 949.5841 & 106 & 5.0 & -10.6\, & \,2.5 & 0.975\\
\hline
\end{tabular}
\end{table}
% Note: possible to add RVs and other indicators measured for this study

%2017/04/18 & 862.7525 & 130 & 4.5 & 3.2$\pm$8.2 & 0.008\\
%2017/04/28 & 872.7921 & 154 & 3.8 & -27.7$\pm$6.6 & 0.120\\
%2017/05/04 & 878.7772 & 157 & 3.7 & -73.5$\pm$6.9 & 0.187\\
%2017/05/13 & 887.8047 & 78 & 7.0 & -36.7$\pm$18.8 & 0.287\\
%2017/05/15 & 889.8118 & 91 & 6.4 & -52.0$\pm$15.1 & 0.310\\
%2017/06/18 & 923.6802 & 105 & 5.4 & 5.8\,\,$\pm$12.5 & 0.687\\
%2017/06/29 & 934.5817 & 96 & 5.7 & 12.7\,$\pm$14.9 & 0.808\\
%2017/07/02 & 937.6045 & 145 & 4.0 & 5.2\,\,$\pm$8.0 & 0.842\\
%2017/07/09 & 944.6375 & 165 & 3.5 & 19.4\,$\pm$6.7 & 0.920\\
%2017/07/14 & 949.5841 & 106 & 5.0 & -10.6$\pm$12.5 & 0.975\\

Spectropolarimetric observations of Proxima Cen were collected using the spectropolarimetric mode of the High Accuracy Radial velocity Planet Searcher velocimeter \citep[HARPS-POL;][resolving power: $\sim$115\,000, wavelength domain: 378-691~nm]{mayor2003,snik2011}. From April to July 2017, 10 sequences of circular polarisation spectra were obtained (ESO program 099.C-0334(A), PI: H\'ebrard), each sequence consisting of four individual spectra taken in different configurations of the retarder. The data reduction is carried out using ESPaDOnS's nominal data reduction pipeline \textsc{Libre-ESpRIT} \citep[e.g.,][]{donati2006b}, inspired by the reduction process described in \citet{donati1997}, and adapted to HARPS-Pol spectra \citep[see][]{hebrard2016}. The four sub-exposures within each spectropolarimeteric sequence are combined to extract Stokes $I$ (unpolarized) and Stokes $V$ (circularly-polarized) spectra in a way to remove systematics and correct for spurious polarization signatures to the first order in circular polarization spectra, \citep{donati1997}. The peak signal-to-noise ratios (S/N) per pixel (0.85~\kms\ velocity bin) of the extracted spectra range from 78 to 165, with a median value of 118. The full journal of observations is given in Table~\ref{tab:obs_prop}.

\begin{table}
    \centering
    \caption{Stellar parameters of Proxima Cen used in this study.}
    \begin{tabular}{c|c|c}
    \hline
    Parameter & Value & Reference \\
    \hline
     Spectral type   & M5.5 & \cite{bessel1991} \\
     Distance     & 1.3012\,$\pm$\,0.0003 pc & \cite{gaia2018} \\
     $T_{\rm{eff}}$ & 2980\,$\pm$\,80 K & \cite{ribas2017} \\
     $\log g$ & 5.02\,$\pm$\,0.18 & \cite{passegger2016}  \\
     \mass\  &  0.120\,$\pm$\,0.003 \msun & \cite{ribas2017} \\
     \rs\    & 0.146\,$\pm$\,0.007 \rsun & \cite{ribas2017} \\
     \avbi   &  600\,$\pm$\,150 $G$ (3$\sigma$) & \cite{reiners2008} \\
     $i$ & 47 $\pm$ 7$^{\circ}$ & This work \\
     \pr & 89.8 $\pm$ 4.0 d & This work \\
    \vs\    &  0.06\,$\pm$\,0.01 \kms & This work \\
     $Ro$  &   0.63$^{+0.14}_{-0.10}$ & \cite{wright2018} \\
     \hline
    \end{tabular}
    \label{tab:star_prop}
\end{table}

%, at ESO's 3.6-m telescope at La Silla, Chile.
%Those signatures are detected by analysing the spectra obtained by combining sub-exposures corresponding to identical azimuths of the HARPS-POL's quater-wave plate (called NULL spectra in the following). 

We apply Least-Squares Deconvolution \citep[LSD,][]{donati1997} to compute average Stokes $I$ and Stokes $V$ profiles for our 10 spectra. This is done using a mask of atomic lines computed from an \textsc{atlas9} local thermodynamical equilibrium model assuming an effective temperature of 3000\,K and a surface gravity of 5.0 \citep{kurucz1993}. The final mask contains $\sim$4000 moderate-to-strong lines \citep[i.e., with depth larger than 40\% as advocated in][]{donati1997} covering the entire HARPS domain. The resulting LSD profiles feature a central wavelength of $625$\,nm, an effective Land\'e factor of 1.25 and a mean relative depth with respect to the continuum of 0.686. Zeeman signatures of full-amplitude up to 0.3\% of the unpolarized continuum level are most of the time detected in the Stokes $V$ spectra. We observe a modulation of the Zeeman signatures, with a sign switch occurring at mid-time throughout the run. In order to ensure that the LSD profiles are not affected by the high level of noise in the bluest part of the spectra (S/N typically of 10-20 for the orders centered on wavelengths lower than 400\,nm), we also extracted the LSD profiles using only the mask lines redder than 400 nm. As the resulting line profiles and magnetic analysis were marginally impacted by the process, we kept using the LSD profiles computed using the full atomic line mask.

%with the stellar properties listed in Table~\ref{tab:star_prop}

% in the stellar rest frame

%Spurious signatures are systematically detected in the spectra obtained by combining sub-exposures corresponding to identical azimuths of the HARPS-POL's quater-wave plate (called NULL spectra in the following). If not corrected, these signals will be interpreted as signatures of toroidal magnetic field, resulting in erroneous magnetic analysis. We thus filter those by modelling the median NULL LSD profile using a Gaussian function whose amplitude is linearly adjusted to match the spurious signatures in each NULL LSD profile, and subtract the best model from the corresponding Stokes $V$ LSD profile.

\section{Magnetic analysis}\label{sec:section3}

\subsection{Longitudinal field}\label{sec:Section3.1}

%\begin{eqnarray}
%B_{\rm{l}} [G] = -2.4 \times 10^{11} \frac{\int v V(v) \rm{d}v }{ \lambda_{0} g_{\rm{eff}} c  \int \left( I_{\rm{C}} - I(v)  \right)dv},
%\label{eq:blong}
%\end{eqnarray}
%\noindent

\begin{figure}
    \centering
    \includegraphics[width=\linewidth]{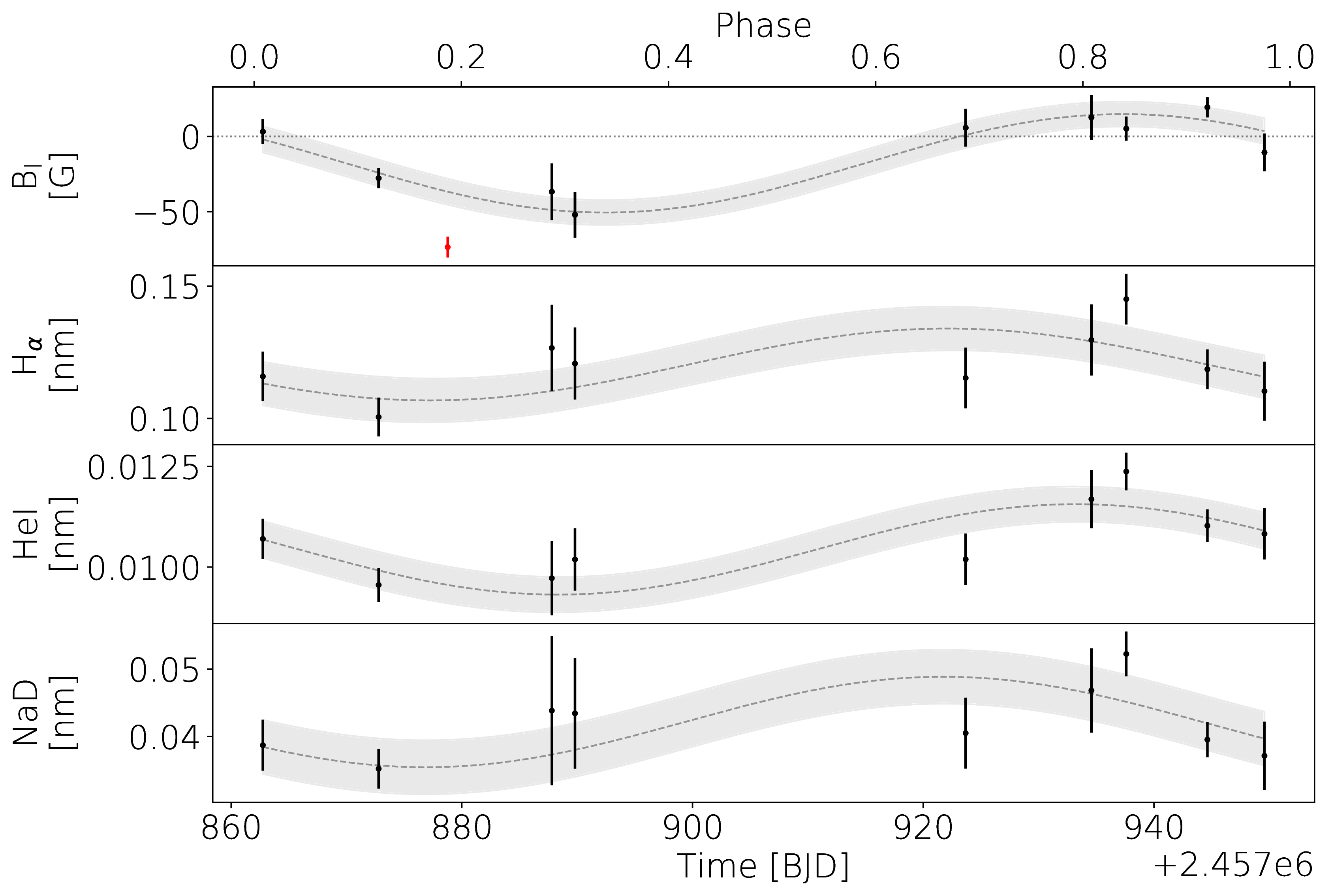}
    \caption{From top to bottom: time-series of the longitudinal field \bl, and H$\alpha$ He\,I and Na\,D equivalent widths. In each panel, we show the data points as black dots (with $\pm$1$\sigma$ error bars), and a simple sine-wave linearly fitted to the data points using a least-squares estimator assuming \pr\,$=$\,89.8\,d in gray dashed lines. Error bars from photon and detector noise on the equivalent widths of the activity proxies (3 lower panels) were respectively scaled up by a factor of about 100, 5 and 10 to tentatively account for their intrinsic variability. In each panel, the gray bands indicate the $\pm$\,1$\sigma$ error bands of the fit. In the top panel, the horizontal dotted line indicates the zero level. The chromospheric indices are 35\% to 85\% larger than the median values of the time-series at phase 0.187, which is interpreted by a stellar flare occurring shortly before the observation. This measurement was not included in the sine-wave fit to the chromospheric indices and is not displayed here for clarity purposes. As a precautionary measure, we also discarded the \bl\ value at phase 0.187 in the fitting procedure, even though it only marginally impacts the results of the fit.}
    \label{fig:fit_bl}
\end{figure}

% Error bars from photon and detector noise on the equivalent widths of the activity proxies (3 lower panels) were respectively scaled up by a factor of xx, yy and zz to tentatively account for their intrinsic variability

%To account for excess of uncorrelated error (due to stellar variability or uncorrected systematics), the error bands are empirically rescaled by the reduced chi-square of the residuasl of the sine-wave fit.}

%The temporal fluctuation of the projected longitudinal magnetic field, \bl, is known to be excellent an proxy of the stellar rotation period \citep{donati2006,hebrard2016}. 

For each observation, we compute the line-of-sight projection of the magnetic field, \bl, by integrating our Stokes $I$ and $V$ LSD profiles on a 23\,\kms-wide window\footnote{The integration window is chosen to include the full width of the observed circularly-polarized Zeeman signatures (see the left panel of Fig.~\ref{fig:stokesV}).} using the method detailed in \citet{donati1997}. \bl\ is a simple proxy of the field geometry, and its temporal fluctuations are known to provide reliable information on the stellar rotation period, \pr\ \citep{donati2006,hebrard2016}. The longitudinal field values, are given in Table~\ref{tab:obs_prop} and range typically from -73 to 20~G with a median 1$\sigma$-uncertainty of 10\,G. We model the \bl\ time-series using a simple sine-wave and use $\chi^{2}$~statistics to estimate \pr. We find that the \bl\ time-series is modulated at \pr\,=\,90\,$\pm$\,6\,d. The best fit to the \bl\ time-series is shown in the top panel of Fig.~\ref{fig:fit_bl} (reaching a reduced-$\chi^{2}$, denoted hereafter \crr, of 1.0).

\subsection{Magnetic reconstruction of Proxima Cen}\label{sec:Section3.2}

\begin{figure}
    \centering
    \includegraphics[width=\linewidth]{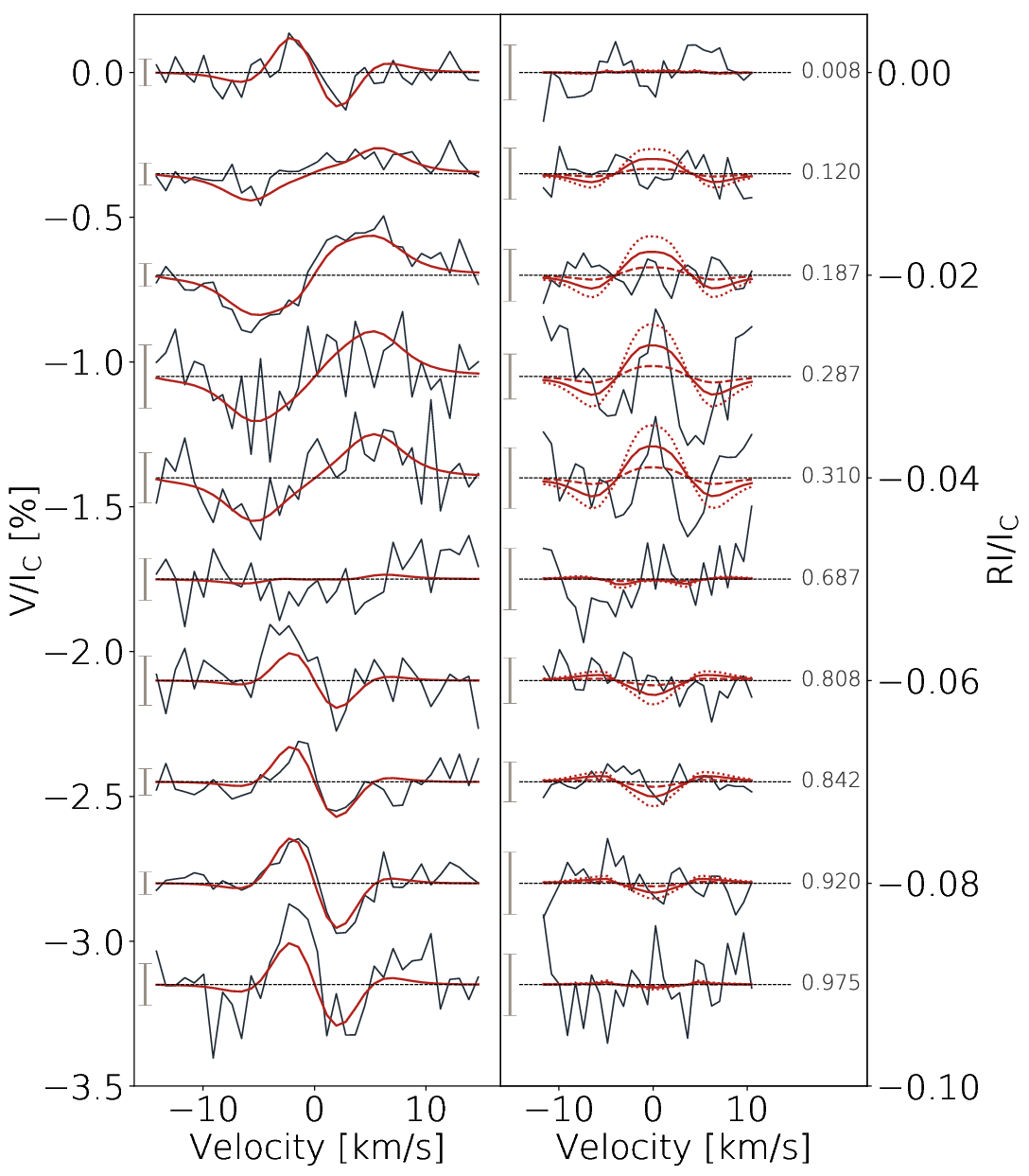}
    \caption{\textit{Left panel:} Time-series of the observed circularly-polarized Zeeman signatures of Proxima Cen (black lines) and maximum-entropy reconstruction (red lines) in the stellar rest frame. \textit{Right panel:} Median-subtracted Stokes $I$ LSD profiles (RI; black solid lines) and predictions using the magnetic topology presented in Sec.~\ref{sec:Section3.2}, assuming \fii/\fv\ of 1 (red dashed lines), 3 (red solid lines) and 5 (red dotted lines, see the text for a definition of \fv\ and \fii). The $\pm$1$\sigma$ error bars on each LSD line are indicated on the left side of each profile in both panels, while the corresponding stellar rotation phase, defined assuming \pr\,=\,89.8\,d and the initial date BJD\,=\,2457862.0,} are written on the right side of the right panel.
    \label{fig:stokesV}
\end{figure}

%the rotation cycle of the star is defined assuming \pr\,=\,89.8\,d and the initial date BJD\,=\,2457862

%To crop your output file with python
%pdfcrop --margins '0 0 0 -205' output.png test.png

\begin{figure*}
    \centering
    \includegraphics[width=\linewidth]{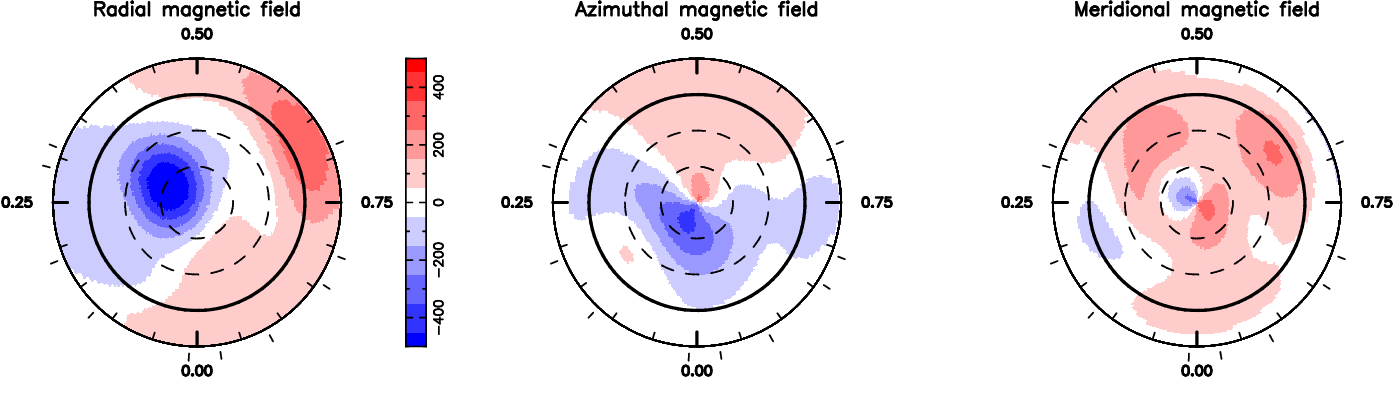}
    \caption{Surface distribution of the radial (left panel), azimuthal (middle panel), and meridional (right panel) components of the large-scale magnetic field of Proxima Cen. The star is described as a flattened polar view, where the circles indicate the equator (solid line) and -30$^{\circ}$, 30$^{\circ}$, and 60$^{\circ}$ parallels (dashed lines). Ticks around the star mark the observations. Magnetic fields are expressed in G. Note that each cell features a filling factor of \fv\,=\,0.1.}
    \label{fig:magmap}
\end{figure*}

%\begin{figure}
%    \centering
%    \includegraphics[width=\linewidth]{mag_coeff.png}
%    \caption{Distribution of the magnetic flux between the coefficients (degree l and order m) of the spherical harmonic expansion of radial (left panel), azimuthal (middle panel) and meridional (right panel) components of the magnetic field vector.}
%    \label{fig:mag_coeff}
%\end{figure}

We use ZDI \citep[][]{semel1989,brown1991,donati1997b} to invert the Stokes $V$ LSD profiles into a map of the large-scale magnetic field at the surface of Proxima Cen. ZDI exploits the rotational modulation of Zeeman signatures to retrieve the poloidal and toroidal components of the large-scale magnetic field, both expressed as weighted sums of spherical harmonics \citep[][]{donati2006}. For a given magnetic map, the stellar surface is sampled into a dense grid of $\sim$10\,000 cells, for each of which ZDI computes local Stokes $I$ and $V$ profiles using analytical expressions from the Unno-Rachkovski's solution of the radiative transfer equation, assuming a plane-parallel Milne-Eddington atmosphere \citep{unno1956}. The width and depth of the local profiles are chosen in a way to minimize the \crr\ between the synthetic intrinsic profile and the median observed one. The local profiles are Doppler-shifted to the projected rotational velocity and weighted depending on limb darkening \citep[assuming a linear law of coefficient 0.8;][]{claret2012}, stellar inclination and local brightness, before being combined together into global Stokes $I$ and $V$ profiles. The time-series of the synthetic profiles are iteratively compared to the observed LSD profiles until reaching the maximum entropy solution for a given level of \crr.

%The rotation period and radius of Proxima Cen indicate that its projected rotational velocity, \vs, must be lower than 0.1~\kms,ruling out any possibility of Doppler imaging and limiting the spherical harmonic expansion to $l \approx 5$ for the magnetic field vector \citep{morin2010}. Assuming a conservative value of \vs~=~0.1~\kms\ for Proxima Cen,

Using ZDI, we perform magnetic reconstructions of the stellar surface at a given level of entropy for different values of the stellar rotation period \pr. By fitting a paraboloid to the resulting distribution of \chs, we find \pr\,=\,89.8\,$\pm$\,4.0~d, in good agreement with the value of the rotation period obtained from the \bl\ time-series (see Sec.~\ref{sec:Section3.1}). A similar process allows to find a stellar inclination of $i$\,=\,47\,$\pm$\,7$^{\circ}$ for Proxima Cen. In what follows, the phase of the star is defined assuming \pr\,=\,89.8\,d and the initial date BJD\,=\,2457862, corresponding to the beginning of our observations. From \pr, $i$ and the stellar radius \rs, we find a projected rotational velocity, \vs, of 0.06\,$\pm$\,0.01~\kms\ and use this value in the ZDI reconstruction. Given the low \vs\ of Proxima Cen, we limit the spherical harmonic expansion to a degree $\ell$\,=\,5. Similarly to \citet{morin2008}, we introduce a filling factor \fv\ defined as the fraction of each cell (constant over the stellar surface) that contributes to polarized Zeeman signatures, yielding an integrated flux of the large-scale field equal to $B_{\rm{V}}$ over the cell. We find that \fv\,$=$\,0.1 minimizes the \crr\ of the fit to both Stokes $V$ and Stokes $I$ LSD profiles (see Sec.~\ref{sec:section3.3}).

The best fit to the observed Stokes $V$ LSD profiles is shown in the left panel of Fig.~\ref{fig:stokesV}. We reach a \crr\ of 1.55 from an initial value of 3.7. The Stokes $V$ time series cannot be fit to a lower \crr\ level, possibly due to the intrinsic variability of the large-scale magnetic field over our observing run or to a slight under-estimation of the assumed error bars. The resulting magnetic topology is shown in Fig.~\ref{fig:magmap}. The average reconstructed large-scale magnetic field is \avbv\,=\,$B$\,$\times$\,\fv\,=\,200\,G, meaning that $B$, the typical field strength present on \fv\,$=$\,10\% of the stellar surface, is equal to $\sim$2\,kG. The magnetic topology is mainly poloidal (92\% of the magnetic energy budget), with a dominant dipolar contribution (60\% of the poloidal energy) featuring a dipole of 135\,G, tilted at 51$^{\circ}$ to the rotation axis towards phase 0.28. We also note a significant quadripolar component (22\% of the poloidal energy), confirming the complex topology of the large-scale field. Moreover, we find a moderate fraction of magnetic energy in axisymmetric modes, \faxi\,$=$\,44\%. Finally, no information on the stellar differential rotation can be inferred from the Stokes $V$ spectra, as the total time-span of the observations barely covers a single stellar rotation.

%\textbf{as evidenced by the increase in amplitude of the Zeeman signatures between phases 0.008 and 0.975}

%\setlength{\tabcolsep}{4.5pt}
%\renewcommand{\arraystretch}{0.1}
%\begin{table}
%    \caption{Main results of the magnetic analysis of Proxima Cen. Columns 1-3 indicate respectively the stellar mass, rotation period (contrained from Stokes $V$ spectra) and Rossby number, $Ro$ \citep{wright2018}. Columns 4 and 5 show the filling factor and average magnetic flux of the ZDI reconstruction. The last 5 columns indicate, from left to right, the fraction of the magnefic energy lying in poloidal (\fpol), dipolar (\fdip), quadripolar (\fquad), octupolar (\foct) and axisymmetric modes (\faxi).}
%    \label{tab:mag_res}
%    \centering
%    \begin{tabular}{cccccccccc}
%    \hline
%      \fv & \avbv\ & \fpol\ & \fdip\ & \fquad\ & \foct\ & \faxi \\
%       & [G] & & & & & \\
%    \hline
%     0.1 & 165 & 0.92 & 0.55 & 0.21 & 0.13 & 0.34 \\
%    \hline
%    \end{tabular}
%\end{table}

\subsection{Zeeman broadening in unpolarized line profiles}\label{sec:section3.3}

%\begin{figure}
%    \centering
%    \includegraphics[width=\linewidth]{fit_broad.png}
%    \caption{Time-series of median-subtracted Stokes $I$ LSD profiles (black solid lines) and predictions using the maximum entropy magnetic topology presented in Section~\ref{sec:Section3.2} (red solid lines), assuming \fii=\fv\ (left panel) and \fii=3*\fv\ (right panel). Similarly to Fig.~\ref{fig:stokesV}, we indicate the rotation cycle and $\pm$1$\sigma$ error bar on both sides of each profiles.}
%    \label{fig:broad_I}
%\end{figure}
%The extracted Stokes $I$ LSD lines exhibit clear variations in width likely due to modulations in the Zeeman broadening with stellar rotation.

We observe a modulation of Zeeman broadening in the Stokes $I$ LSD profiles. In particular, the profiles appear significantly broadened around phase 0.3, when the magnetic pole points towards the observer. We generate a time-series of synthetic Stokes $I$ profiles from the magnetic topology obtained in Sec.~\ref{sec:Section3.2}. As for the modeling of Stokes $V$ spectra, we introduce a filling factor \fii, also constant over the stellar surface and defined as the fractional area of each cell containing a small-scale magnetic field of strength $B_{\rm{I}}$\,=\,$B_{\rm{V}}$/$f_{\rm{V}}$, i.e., assumed to be distributed as the large-scale field $B_{\rm{V}}$ \citep[see][for a more extensive definition of \fii]{morin2008}. The difference between \fii\ and \fv\ is explained by the fact that circular polarization signatures and Zeeman broadening probe different scales of the magnetic field. On small scales, regions with opposite magnetic polarities produce circularly-polarized signals that almost completely cancel out, which is not the case for the Zeeman broadening in Stokes $I$ spectra. We compare the Stokes $I$ LSD profiles to the synthetic ones assuming various values for \fii\ and find that \fii\,$\sim$\,3\fv\ minimizes the \chs\ of the residuals in the central regions of the Stokes $I$ profiles (i.e., within $\pm$6\,\kms\ of the line center) where the modulation of the Zeeman broadening is largest (see the left panel of Fig.~\ref{fig:stokesV}). We however caution that, being based on the simplifying approximation that $B_{\rm{I}}$ is distributed as $B_{\rm{V}}$, our measurement of \fii/\fv\ is likely no more than a rough estimate. Using \fii\,=\,3\fv, we find an unsigned magnetic flux density \avbi\,\fii\,=\,\avbv\,\fii/\fv\,$\sim$\,600\,G, consistent with the flux density measured from the Zeeman broadening of unsigned lines in \citet{reiners2008}. The ratio \fv/\fii\,$\sim$\,0.33 is apparently higher than that typically measured for FC \mdw s when this ratio was shown to reach 0.15 when the dynamo is saturated \citep[][]{reiners2009,morin2010,see2019}. The reasonably good agreement between the large- and small-scale magnetic topologies and the high ratio \fv/\fii\ suggest that the underlying dynamo process is apparently efficient at injecting the magnetic field into the largest spatial scales. Finally, assuming larger values for \fv\ in the magnetic reconstruction yields poorer fits to the modulation of the Zeeman broadening in the Stokes $I$ LSD profiles, confirming that \fv\,=\,0.1 is a good estimate of the filling factor (with a typical uncertainty of 0.03).

\begin{figure*}
    \centering
    \includegraphics[width=\linewidth]{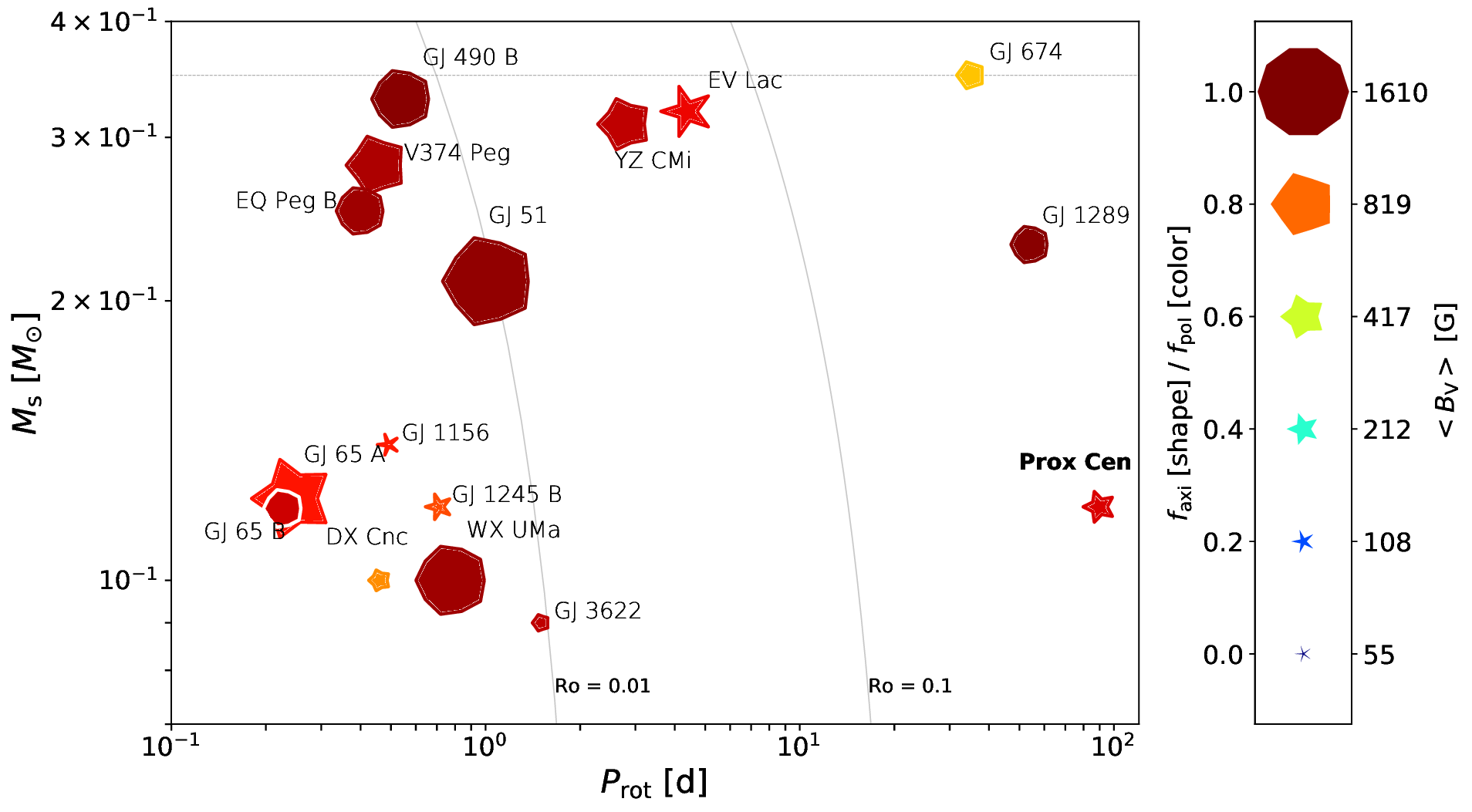}
    \caption{Rotation period-mass diagram of FC \mdw s with reconstructed magnetic topology. The size of the symbols is proportional the strength of the magnetic field, their shapes represent their degree of axisymmetry (decagons and stars with \faxi\ of 1 and 0 respectively), and their colors refer to the fraction of poloidal field of the reconstruction (blue and red standing respectively for pure toroidal and poloidal fields). Contours of constant Rossby numbers of 0.1 and 0.01, computed using \citet{wright2018} empirical relationship, are plotted in gray solid lines. The horizontal line indicate the theoretical full-convection limit \citep[$\sim0.35$\msun][]{baraffe1998}. Except for Proxima Cen, the data originate from the magnetic analyses presented in \citet{morin2008,phan-bao2009,morin2010,morin2011b,kochukov2017,moutou2017}.}
    \label{fig:confusiogram}
\end{figure*}

\section{Discussion and conclusion}\label{sec:section4}

In this study, we report the first detection of a large scale magnetic field at the surface of Proxima Centauri. Ten spectropolarimetric observations, obtained with HARPS-Pol from April to July 2017, exhibit Zeeman signatures whose modulation indicates a stellar rotation period of 89.8$\pm$4.0\,d and suggests a stellar inclination of 47$\pm$7$^{\circ}$. We find a large-scale magnetic field of $\sim$200\,G, whose topology is mainly poloidal, and features a dipole component of 135\,G tilted at 51$^{\circ}$ to the rotation axis. Due to the low \vs\ of the star, we mostly access the largest scales of the magnetic field (i.e., $\ell$\,$\leq$\,5). Our resulting magnetic topology is thus expected to be no more than weakly affected by the gap in our observations between phases 0.310 and 0.687. The circularly-polarized magnetic flux density is only $\sim$3$\times$ smaller than the flux measured from the Zeeman broadening of unsigned lines, which suggests that the underlying dynamo process is efficient at injecting magnetic energy in the largest spatial scales. As illustrated in the rotation period-mass diagram shown in Fig.~\ref{fig:confusiogram}, Proxima Cen's magnetic properties appear similar in term of field strength, fraction of poloidal energy, and degree of axisymmetry, to those of the group of FC \mdw s with saturated dynamo exhibiting multipolar large-scale fields \citep{morin2010}. However, Proxima Cen has the highest ratio \avbv/\avbi\ reported so far, whereas this ratio is generally significantly weaker for stars in the multipolar regime ($\sim$6\%) than for stars in the dipole dominated group ($\sim$15\%). These results, if confirmed by new spectropolarimetric observations of Proxima Cen and of other slowly-rotating FC stars, challenge our understanding of dynamo generation in \mdw s. We now discuss the implications of the magnetic analysis on the chromospheric activity and magnetic cycle of the star, and on its extended magnetospheric structure.

%\textbf{The significant gap between phases 0.310 and 0.687 in our observations is not expected to have a significant impact on the recovered dipole, well constrained from the observed Zeeman signatures. However, this could slightly affect the position of the magnetic equator, barely constrained from our phase sampling, and higher order magnetic components recovered with ZDI (e.g., quadripolar and octupolar fields) due to potential unobserved structures in the phase gap.}

% \subsection{Comparison to rapidly rotating low-mass stars}

\subsection{Magnetic activity and cycle of Proxima Cen}

%The large-scale magnetic topology recovered with ZDI gives valuable information about the distribution of active regions at the surface photosphere. 

Optical chromospheric activity indicators are known to be well-coupled to bright and dark features in the case of the Sun. Comparing their temporal evolution with the magnetic topology recovered with ZDI provides us with valuable information on the magnetic connections between the photosphere and the chromosphere. We compute the equivalent width of H$\alpha$ (6562.808\,\angstrom), He\,I D3 (5875.62\,\angstrom), and Na\,I D1 and D2 (resp. 5895.92 and 5889.95\,\angstrom) chromospheric lines\footnote{Note that, due to the very low S/N in the bluest parts of the spectra, we could not reliably estimate the equivalent width of Ca H\,\&\,K lines from our observations.}. The integration windows are those defined in \citet{gomes2011} for late-type stars. 

%Error bars from photon and detector noise on the equivalent widths of the activity proxies (in the 

% For each line, the error bars are estimated from the dispersion in the reference continuum bands of \citet{gomes2011}. 

%This is done by computing the flux in each line and dividing it by the continuum in reference bands at both sides of the line.

%Error bars from photon and detector noise on the equivalent widths of the activity proxies (3 lower panels) were respectively scaled up by a factor of xx, yy and zz to tentatively account for their intrinsic variability'

%Ca\,II H \& K (resp. 3968.47 and 3933.66\,\angstrom; also known as Mount Wilson S index)

As shown in Fig.~\ref{fig:fit_bl}, the time-series of chromospheric activity indicators appear modulated at \pr. The chromospheric emission in the He\,I line varies in phase with the magnetic field, reaching its minimum at phase 0.29\,$\pm$\,0.06, when the negative pole comes closest to the observer (i.e., at phase 0.34\,$\pm$\,0.03), and its maximum when the magnetic equator gets closer to the line-of-sight (phase $\sim$0.8). The emissions in the cores of H$\alpha$ and Na\,D lines are shifted by $\sim$0.2 in phase with respect to the magnetic field, with minimums of emission respectively reached at phases 0.17\,$\pm$\,0.07 and 0.17\,$\pm$\,0.08 (consistent at 2$\sigma$ with the phasing of the magnetic field). This shift is most likely attributable to (i)~a complex distribution of the chromospheric material which might be differently probed in He\,I and H$\alpha$/Na\,D lines or (ii)~to intrinsic stellar variability (e.g., flares or slow temporal evolution of the stellar chromosphere). The observed modulation of the chromospheric indices may reflect the coronal hole (associated with open field lines and presumably darkest at the chromospheric level) going in and out of view as the star rotates, and being best and least visible at phases 0.3 and 0.8 respectively. A full radiative transfer computation would be needed to confirm this suggestion.

%which remains consistent at 2$\sigma$ with the magnetic field,

%\textbf{The emissions in the cores of H$\alpha$ and Na\,D lines are shifted in phase with respect to the magnetic field, with minimums of emission respectively reached at phases 0.12\,$\pm$\,0.15 and 0.10\,$\pm$\,0.07. The reason for this phase shift is unclear but likely attributable to different sensitivity of each indicator to stellar variability and to the scarcity of our data set.}

% \textbf{We note that the curve of S index is significantly shifted compared to that of the other activity indicators. We interpret this observation as resulting from the strong sensitivity of Proxima Cen's S index to stellar variability \citep[e.g.,][]{gomes2011} and the low S/N in the continuum regions used in the computation of this indicator.}

% \footnote{This difference in phase between the magnetic topology and chromospheric emission suggests that the position of large-scale magnetic dipole recovered with ZDI is slightly erroneous (by $\sim$0.1 in phase), probably due to the gap between phases 0.310 and 0.687 in our observations. However, we caution that this may just be the result of stellar variability affecting the values of the chromospheric activity indices.}.}

Our spectropolarimetric observations take place near Proxima Cen's activity maximum \citep{suarez2016,wargelin2017}. If, as suggested in the literature, Proxima Cen undergoes a 7\,yr-solar-like activity cycle powered by a $\alpha \Omega$-dynamo process \citep[e.g.,][]{yadav2016}, one would expect the magnetic field to oscillate from a nearly axisymmetric dipole, at the activity minimum, to a weaker less axisymmetric multipolar field, at activity maximum \citep[][]{kitchatinov2014}. In particular, the total fractions of axisymmetric and poloidal magnetic energies reconstructed by ZDI seem to be excellent proxies of the magnetic cycle for partly-convective stars (see Lehmann et al., 2020, in prep.). Hence the interest to keep monitoring Proxima Cen's magnetic properties throughout the activity cycle.

Another major interest of future spectropolarimetric observations is to constrain the stellar differential rotation and its evolution with the activity cycle. Our measurement of \pr, although significantly higher than the average photometric rotation period \citep[e.g., 83.10$\pm$0.05\,d in][]{wargelin2017}, is consistent with the period of 90.1\,d measured by \citet{wargelin2017} in 2009, i.e., around the previous activity maximum. The authors do not find any correlation between the photometric \pr\ and the cycle phase, which suggests that the photometric period is not so good proxy of the activity cycle for proxima Cen. On the other hand, the variation of \pr\ measured from circularly-polarized Zeeman signatures might be a good indicator of the magnetic cycle, as the large-scale magnetic field is directly linked to the underlying dynamo processes.

Finally, near-infrared spectroscopic observations of Proxima Cen with upcoming instruments like NIRPS \citep{bouchy2017,wildi2017} will complement the analysis by allowing the measurement of other activity indicators (e.g., based on the emission flux in the core of the He\,I triplet at 1083\,nm, and Paschen $\beta$), and by constraining the ratio \avbi/\avbv\ thanks to the larger Zeeman broadening in the near-infrared than in the optical domain.

\subsection{Extended magnetosphere and implications for Proxima-b}

\begin{figure}
    \centering
    \includegraphics[width=\linewidth]{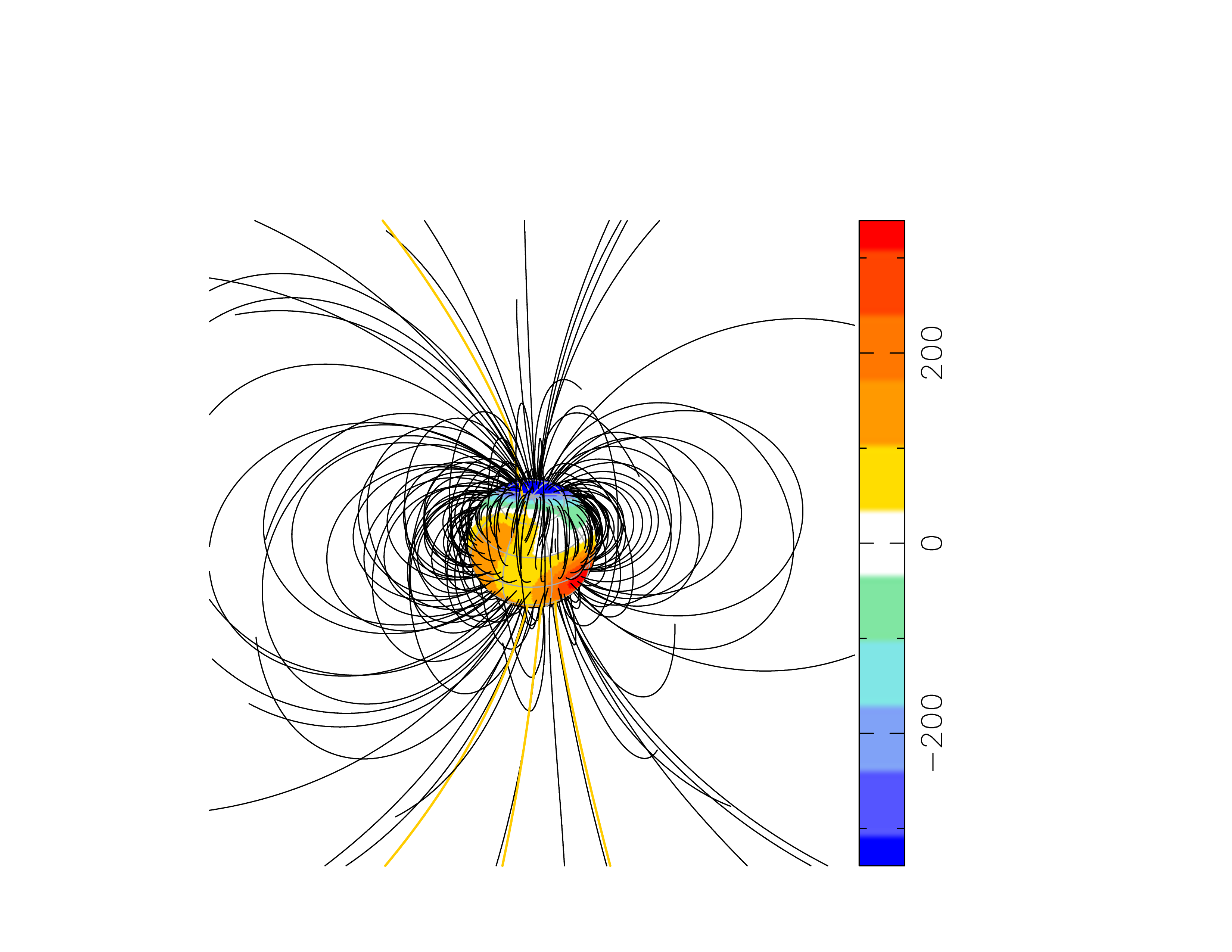}
    \caption{Potential field extrapolated from the reconstructed magnetic topology at phase 0.8 (around the maximum H$\alpha$ emission). Open and close field lines are respectively shown in black and yellow solid lines. The color scale depicts the strength of the radial component of the large-scale magnetic field at the stellar surface. In order to better visualize the variations of the radial field at the surface of the star (and then more directly compare it with the potential field extrapolation), we show the star as seen from a distance of 5\,\rs. The source surface at which the magnetic field lines are radial is set to 25\,\rs\ for this plot.}
    \label{fig:extrap}
\end{figure}

Given that the corotation radius of proxima Cen is $\approx$\,285\,\rs, field lines are not expected to break under the effect of centrifugal forces, but rather under that of the stellar wind. We estimate the typical radius of the spherical Alv\`en surface, \ra, above which field lines open under stellar wind ram pressure. We first compute the magnetic confinement parameter, $\eta$, defined in \citet{uddoula} as the ratio between magnetic and wind kinetic energy densities, assuming a magnetic field of 135\,G. Using a stellar mass-loss rate of 2$\times$\,10$^{-15}$\,\msun/yr and a stellar wind terminal velocity of 400~\kms\ \citep{wood2001}, we obtain a confinement parameter as large as $\eta$\,$\sim$\,$3.9 \times 10^{5}$. At large distance from the star (typically a few stellar radii), the extended magnetic field of Proxima Cen is mainly dipolar. Moreover, Proxima Cen features a corona likely reaching the MK level, which would be enough to entirely ionize the stellar wind \citep[as demonstrated for a few \mdw s of earlier types; e.g.,][]{vidotto2019,mesquita2020}. Under these conditions, \citet{uddoula} semi-analytical relations provide us with a first order estimate of the equatorial Aflv\`en radius of $\sim$25\,\rs. At this distance, the magnetic field of Proxima Cen is mostly dipolar, as evidenced by the extrapolation of the potential field lines shown at phase 0.8 in Fig.~\ref{fig:extrap}, computed using the field-extrapolation technique described in \citet{jardine1999}. Proxima-b, that orbits at a distance of $\sim$70\,\rs\,$>$\,\ra\ from its host star, is thus expected to lie in the super-Alfvenic regime, where no direct star-planet magnetic connection occurs \citep{strugarek2015}. We however caution that magneto-hydrodynamical simulations are needed to validate this conclusion \citep[e.g., in a way similar to][]{strugarek2015,vidotto2017,folsom2018,strugarek2019,folsom2019}. Finally, constraining the evolution of the size of the magnetosphere with the stellar activity cycle will help unveiling potential SPIs that may affect the magnetosphere of Proxima-b.

As discussed in \citet{vidotto2014b}, the value \faxi\ of the ZDI reconstruction correlates with the flux of galactic cosmic rays received by the planet \citep[based on solar system observations studied in][]{wang2006}. The moderate value of \faxi\ for Proxima Cen indicates that, its close-in planet is relatively well-shielded from galactic cosmic rays at the time of the observations, which has implications for habitability condition studies \citep{rimmer2013,Sadovski2018}. Our analysis confirms the strength of the large-scale field assumed in \citet{ribas2016}, which implies that Proxima b could likely sustain a magnetosphere of $\sim$2-3 planetary radii under a wind terminal velocity similar to that assumed to compute the Alfv\`en radius of the star. Moreover, extended non-axisymmetric magnetic fields tend to favour axisymmetric distributions of the wind mass flux \citep{vidotto2014b}. This suggests that the size of Proxima-b's magnetosphere undergoes little variations at the time of observations, contrary to what was predicted by recent studies based on a different magnetic topology \citep[e.g.,][who assume a Gj\,51-like magnetic geometry whose maximum strength is rescaled to 600\,G]{garraffo2016}.

% value for \faxi\ for Proxima Cen indicates that the close-in planet is likely well-shielded from cosmic rays, which are known to highly affect exoplanet atmospheric and habitability properties \citep[e.g.,][]{rimmer2013}
%First, the value \faxi\ obtained with ZDI, reliable at large distance from the star \citep{vidotto2014b}, correlates with the flux of cosmic rays received by the planet \citep[based on solar observations studied in][]{wang2006}. The relatively low value for \faxi\ for Proxima Cen indicates that the close-in planet is likely well-shielded from cosmic rays, which are known to highly affect exoplanet atmospheric and habitability properties \citep[e.g.,][]{rimmer2013}. Second, non-axisymmetric fields tend to favour axisymmetric distributions of the wind mass flux, implying thus a relatively constant size of the planet magnetosphere over time. These results are not in line with those reported in \citet{garraffo2016}, who assume a a Gj\,51-like magnetic geometry for Proxima Cen, which turns out to be different from the one that we find in this work (see Fig.~\ref{fig:confusiogram}).

%%% General conclusion
More generally, our study confirms that the large-scale magnetic topology of slowly-rotating FC \mdw s 
can be retrieved with ZDI. Our results open up promising prospects for the study of the magnetic field of FC \mdw s and their interactions with close-in planets. In particular, new generation near-infrared spectropolarimeters like SPIRou at the Canada-France-Hawaii Telescope \citep{donati2018} and its upcoming twin, SPIP, at the Pic du Midi Observatory, may bring decisive constraints on the dynamo processes operating at the lower end of the main sequence.

\section*{Acknowledgements}

This project was funded by the European Research Council (ERC) under the H2020 research \& innovation programme (grant agreements \#740651 NewWorlds). This work is based on observations made with ESO Telescopes at the La Silla Paranal Observatory. We thank the referee for valuable comments and suggestions which helped us improving an earlier version of the manuscript.

\section*{Data availability}

This paper includes data collected by the HARPS-Pol spectropolarimeter, which is publicly available from the ESO Science Archive Facility (program ID: 099.C-0334).

\bibliographystyle{mnras}
\bibliography{bibliography} % if your bibtex file is called example.bib

%%%%%%%%%%%%%%%%%%%%%%%%%%%%%%%%%%%%%%%%%%%%%%%%%%

%%%%%%%%%%%%%%%%% APPENDICES %%%%%%%%%%%%%%%%%%%%%

% Don't change these lines
\bsp	% typesetting comment
\label{lastpage}
\end{document}